\documentclass[aps,showpacs,a4paper,floatfix,twocolumn,prl,amsmath,amssymb]{revtex4-1}

\usepackage{xcolor,ulem} 

\usepackage[utf8]{inputenc}
\usepackage[T1]{fontenc}
\usepackage[english]{babel}
\usepackage{lmodern} 

\usepackage{booktabs} 
\usepackage{array} 
\usepackage{multirow} 
\usepackage{tabularx} 
\usepackage{subcaption} 
\usepackage{graphicx} 
\usepackage{subcaption} 

\usepackage{amsmath}
\usepackage{amssymb}
\usepackage{mathtools}
\usepackage{bm} 

\usepackage{graphicx}
\usepackage{epsfig,hyperref}
\usepackage{color} 
\usepackage{morefloats}
\usepackage{rotating}
\usepackage{relsize}
\usepackage{url}
\usepackage{ragged2e}

\usepackage{booktabs}
\usepackage{soul}
\usepackage{array}

\usepackage{import}

\begin{document}

\title{
Calibrating the medium effects of light clusters in heavy-ion collisions
}

\author{Tiago Cust{\'o}dio$^1$}
\author{Alex Rebillard-Souli{\'e}$^2$}
\author{R{\'e}mi Bougault$^2$}
\author{Diego Gruyer$^2$}
\author{Francesca Gulminelli$^2$}
\author{Tuhin Malik$^1$}
\author{Helena Pais$^1$}
\author{Constan\c ca Provid{\^e}ncia$^1$}

\affiliation{$^1$CFisUC, Department of Physics, University of Coimbra,
  3004-516 Coimbra, Portugal. \\
$^2$Normandie Univ., ENSICAEN, UNICAEN, CNRS/IN2P3, LPC Caen, F-14000 Caen, France. }



\date{\today}
\begin{abstract}
We propose a 
Bayesian inference estimation of in-medium modification of the cluster self-energies from light nuclei multiplicities measured in selected samples of central $^{136,124}$Xe$+^{124,112}$Sn collisions with the INDRA apparatus. The data are interpreted with a relativistic quasi-particle cluster approach in the  mean-field approximation without any prior assumption on the thermal parameters of the model. 
An excellent reproduction is obtained for H and He isotope multiplicities, and compatible posterior distributions are found for the unknown thermal parameters. 

We conclude that the cluster-$\sigma$-meson coupling is temperature dependent, becoming weaker when the temperature increases, in agreement with microscopic quantum statistical calculations. This implies a faster decrease of the light cluster abundances with temperature than previously estimated. 

\end{abstract}

\maketitle

\newpage

{\it Introduction:} With the increase of precision on the determination of the static properties and thermal evolution of compact objects by multi-messenger observations, neutron stars can be considered as true laboratories of baryonic matter under extreme conditions. 
In particular, it is expected that the wider frequency window and larger sensitivity of the next generation of gravitational wave (GW) detectors \cite{Branchesi:2023mws,Evans:2021gyd},
will allow the 
detection of supernova and post-merger signals \cite{Chatziioannou:2021tdi, Radice:2020ddv,Carson:2019rjx}.  Larger sensitivity 
and the interpretation of non-modelled GW waveforms with their connection with the electromagnetic counterpart from numerical relativity simulations, require the use of temperature and composition dependent EoS \cite{Nedora_2021}, with in particular the inclusion of light cluster formation \cite{Typel2009}. 
 Their abundance can indeed affect the dynamics and properties of supernovae \cite{Arcones2008,Sumiyoshi:2008qv,Furusawa:2013tta} and binary neutron star mergers \cite{Bauswein:2013yna,Rosswog2015,Radice:2018pdn,Navo:2022xle,Psaltis:2023jvk}, both directly through their weak reactions with the surrounding medium, 
 and indirectly through their competition with heavy nuclei \cite{Pais_2019}, which can modify the proton fraction and the size of nucleosynthesis seeds  \cite{Nedora_2021}.  They  can also  have a significant (indirect) effect on the dynamics of the core-collapse supernova explosion giving rise to a faster shock retreat and an early neutrino luminosity \cite{Fischer_2020},  even though, only a negligible (direct) impact from the weak reactions involving  the  light clusters was obtained. The transport coefficients are determined by the collision rates of electrons and/or neutrinos with clusters, which in turn depend on the cluster abundances and sizes. In binary mergers, the recombination of free nucleons into $\alpha$ particles can generate enough energy to induce mass outflows \cite{Beloborodov:2008nx,Lee:2009uc,Fernandez:2012kh}.

Ab-initio calculations of nuclear clustering in dense media start to be available \cite{Ropke:2011tr,Ropke2015,Ropke_2020,Ren_2024},
but general purpose EoS used in large scale astrophysical simulations \cite{CompOse} still require the use of mean-field models where light clusters are introduced as independent quasi-particles coupled to effective mesonic fields \cite{Typel2009,Pais_2019} with couplings calibrated on heavy-ion (HI) experimental data \cite{Qin2011,indra}. 
In the first attempts of optimizing the cluster couplings on HI data \cite{Qin2011,Hempel2015,indra}, 
statistical isothermal samples were tentatively selected by sorting the data in $N_v=13$ bins of the 
average radial velocity ($v_{\text{surf}}$) in the expanding source reference frame. 
The baryonic densities and temperatures associated with the samples were obtained by considering that the statistical ensembles of particles could be described by an ideal gas of classical clusters in the grand canonical ensemble \cite{Qin2011,Hempel2015,indra}, while in a more recent work \cite{Pais2020,Pais2020prl} a parameterised correction to ideal Boltzmann yields was added. 
In the present study, we propose to analyse the data avoiding the ideal gas assumption, thus providing for the first time hypothesis-independent estimation of the in-medium corrections. 
The compatibility of the results using independent analyses of four different data sets varying in size and isospin content will a-posteriori 
  support the interpretation of the
 statistical character of the samples, and allow estimating the systematical uncertainty within a Bayesian framework.

{\it Data analysis:} 
For this analysis, we use selected INDRA data sets corresponding to the center of mass (CM) emitting source produced with four different entrance channels $^{136,124}$Xe$+^{124,112}$Sn at 32 MeV/nucleon \cite{Bougault_2018}. 
Only central events are considered and data are sorted in bins of the average  Coulomb-corrected particle velocities $v_{\text{surf}}$ in the CM frame \cite{Qin2011,indra},
that was shown to be correlated to the dynamics of the expansion, and therefore to the effective temperature of the source \cite{Qin2011}.

 An experimental indication of chemical equilibrium in the selected samples was reported 
 for all Hydrogen and Helium isotopes with the exception of ${}^{6}\text{He}$ \cite{Rebillard-Soulié_2024} which suffers from finite-size effects and is therefore excluded from the analysis.
The global proton fraction of the samples can be safely deduced from the data with minimal hypotheses on the undetected neutrons \cite{indra}, but other thermodynamic quantities such as the baryon density $\rho$ and the temperature $T$ are not measured experimentally and must be obtained indirectly from the chemical equilibrium requirement. To this aim, we use a relativistic mean-field formalism, which includes nucleons and light nuclear clusters as independent quasi-particles and takes into account the expected in-medium effects of light clusters in a dense medium \cite{Typel2009,Ferreira:2012ha,Pais:2015xoa,Pais2018}.   
 The interactions 
 are mediated by the exchange of virtual mesons: the isoscalar-scalar $\sigma$-meson, the isoscalar-vector $\omega$-meson and  the isovector-vector $\rho$-meson. The corresponding Lagrangian density reads \cite{Pais2018}
\begin{equation}\label{Lagrangian_nlm_inicial}
	\mathcal{L}=\sum_{\substack{j=n,p,\\{}^{2}\text{H},{}^{3}\text{H},\\{}^{3}\text{He},{}^{4}\text{He}}} \mathcal{L}_{j} +\sum_{\substack{m=\sigma,\omega,\rho  }}\mathcal{L}_m + \mathcal{L}_{NL}\,,
\end{equation} 
where the first term describes the nucleons and light clusters interacting with the mesons, the second term describes the meson fields, 
and the last term 
includes non-linear meson terms and is  present in models that do not have density dependent couplings, see \cite{Pais2018,Custodio2020}.
The meson-nucleon couplings, denoted by $g_{mN}$, where $m=\sigma,\, \omega, \, \rho$, are fitted to the nuclear properties. In the following we present results considering the FSU parameterization \cite{Todd-Rutel:2005yzo} 
and the DD2 parameterization with density dependent couplings \cite{Typel2009}, which have been calibrated to the ground state properties of finite nuclei and, in the case of the former, also to their linear response.
Both parameterizations are therefore suitable for the description of low-density matter, 
 and the difference between the two predictions can be taken as an estimation of the model dependence of our results.
 The meson-cluster couplings are defined as $g_{\sigma j}=x_{s} A_j g_{\sigma N}$, 
$g_{\omega j}=A_j\,g_{\omega N}$, where $A_j$ is the mass number of cluster and the 
coupling ratio $x_{s}(\rho,T)$ measures the (possibly density and temperature dependent) in-medium modification of the cluster self-energies, and is calibrated on the experimental data.
 Note that there is some degeneracy between the couplings of the $\sigma$ and $\omega$ mesons to the clusters, as discussed in \cite{Ferreira:2012ha}. In the present study, the coupling to the $\omega$ meson was fixed and the coupling to the $\sigma$ was calibrated to the experimental data. It is expected that this degeneracy will be removed if a complete transport description of the HIC is performed, since the role of each  meson in the dynamics of the system is different.

\begin{figure}[tp]
	\centering
		\includegraphics[width=1.\linewidth]{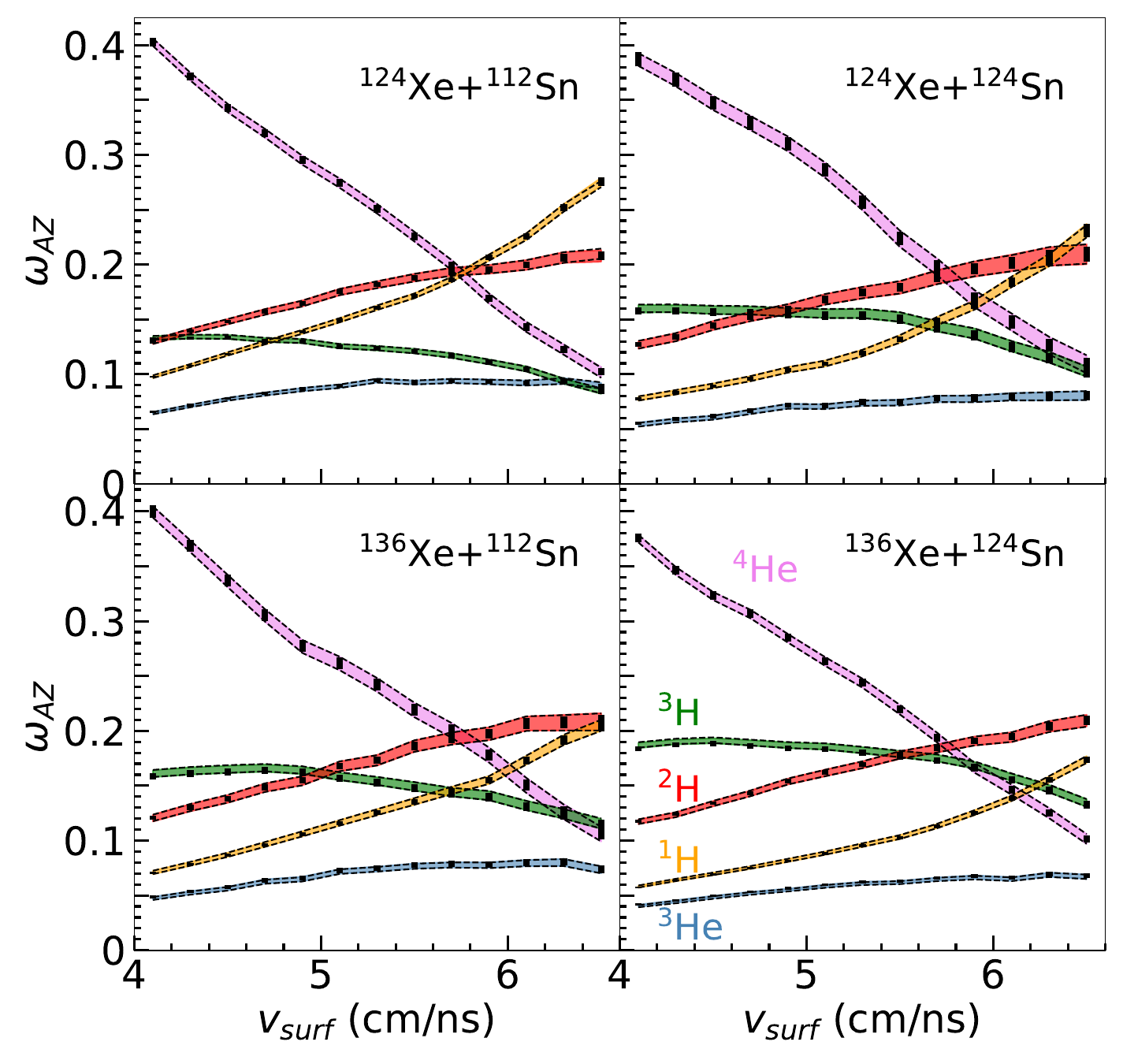}
  \caption{ \justifying Experimental (black symbols) and theoretical (bands) nuclear species mass fractions for the different colliding systems $^{136,124}$Xe$+^{124,112}$Sn and  with the optimised $(\rho, T, x_s)$ parameter distributions displayed in Figs.\ref{fig_rho_T_sets_of_3},\ref{fig_fix_dens_overlap_reg_Tvsxs_13}, for DD2 (filled colour bands) and FSU (bands whose contours are black dashed) RMF models. Both experimental black symbols and theoretical bands are represented with 2-$\sigma$ uncertainty.    DD2 and FSU bands mostly agree within accuracy of the plot.}
	\label{Mass_Fractions_vsurf_groups_of_three}
\end{figure}

\begin{figure}[tp]
	\centering
		\includegraphics[width=1.\linewidth]{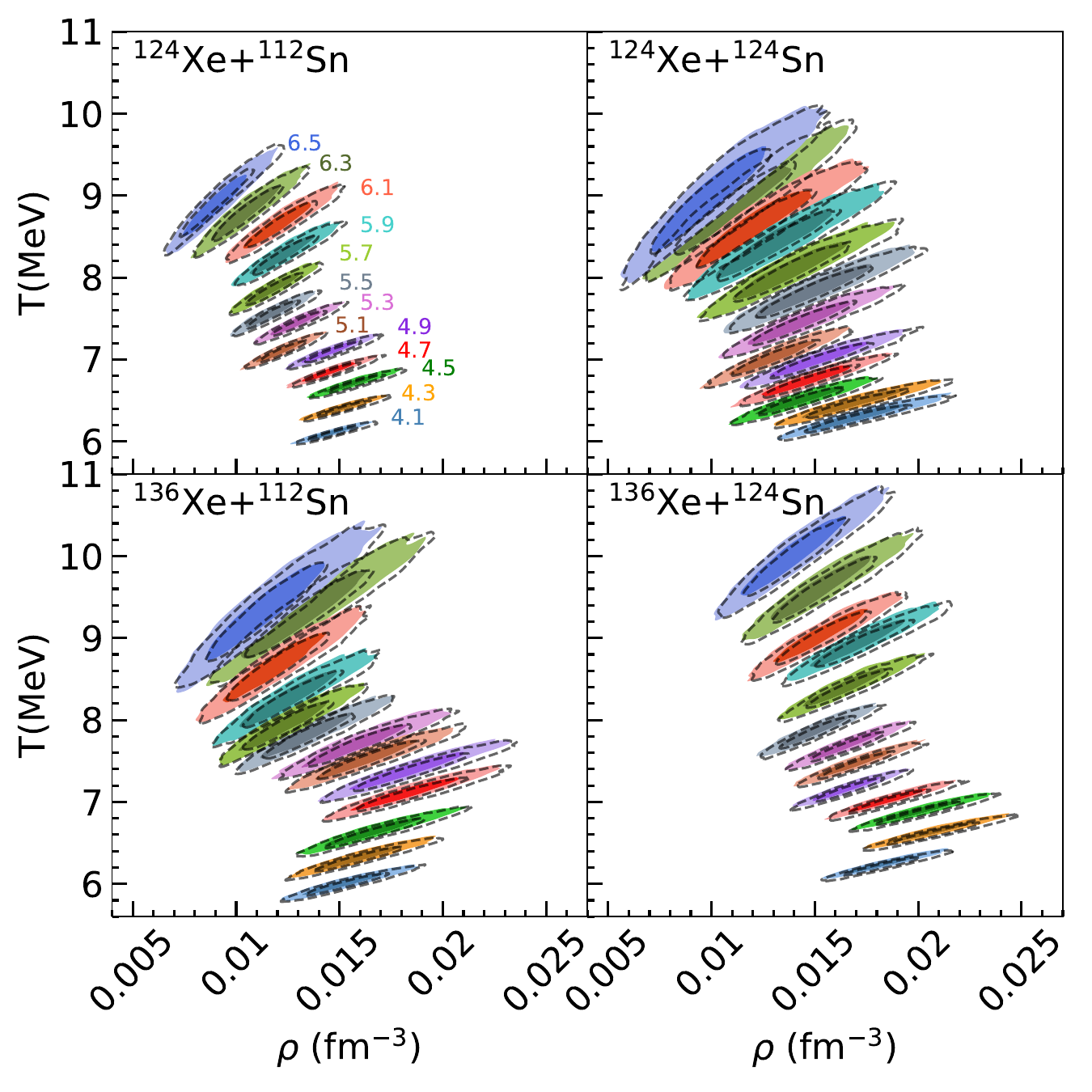}
  \caption{ \justifying Bayesian estimation of the thermodynamical parameters $\rho$ and $T$ for the different experimental samples. 
  Each colour represents a different $v_{\text{surf}}$(cm/ns) bin and is identified on the upper left panel. 
  For each band, the 1,2-$\sigma$ uncertainty regions are shown for DD2 (filled contours) and FSU (unfilled contours delimited with grey dashed lines) RMF models.} 
	\label{fig_rho_T_sets_of_3}
\end{figure}

We carry out $4N_v=52$, one for each system and $v_{\text{surf}}$ bin, independent Bayesian inferences on the measured mass fractions 
$\omega_{AZ}=A Y_{AZ} / A_T$  of the different clusters with mass number $A$ and charge $Z$, where $Y_{AZ}$ are the multiplicities and $A_T$ is the total number of nucleons.  
Independent posterior distributions of the model parameters ${\bf\theta}=\{T,\rho,x_s(\rho,T)\}$ 
are obtained for each $i=1,\dots,4N_v$  velocity bin and each system, according to:
\begin{equation} \label{eq:Bayes}
    p_i \left( {\bf\theta}| \{\omega_{AZ} \}\right )=\frac{p_{{\bf\theta}}}{\cal Z}{\cal L_{\rm g}} \left( \{\omega_{AZ} \}_i|\bf{\theta}\right ),
\end{equation}
where $p_{{\bf\theta}}$ is a flat prior, ${\cal Z}$ is the evidence, 
$\{\omega_{AZ} \}_i$ is the $i-$th mass fraction data set from Ref.\cite{Rebillard-Soulié_2024}, and $\cal L_{\rm g}$ is a gaussian likelihood \cite{Gelman2013}
. 
To perform the inference, we have  used  
the \texttt{PyMultiNest} sampler \cite{Buchner:2014nha,Buchner:2021kpm}, which divides the posterior into nested slices with initial \textit{n-live} points, generating samples from each and recombining them to restore the original distribution \cite{Skilling2004}.

The quality of the calibration is clearly seen from Fig. \ref{Mass_Fractions_vsurf_groups_of_three} where the experimentally measured mass fractions (black symbols) are compared with the  corresponding marginalised posteriors obtained integrating Eq.(\ref{eq:Bayes}) over the $(T,\rho,x_s)$ distributions, 
and does not depend on the model.
This is at variance with the calibration obtained in previous
analyses from the equilibrium constants \cite{Pais2020prl,Pais2020}. It is here verified that the individual $^2$H, $^3$H and $^3$He abundances are not reproduced, and the experimental $^4$He abundances are outside the theoretical predictions for some temperature ranges, while $^1$H is reasonably well reproduced (see Fig. 7 of the Supplemental Material).

This seems to imply that important information is lost when the equilibrium constants, a derived quantity from the directly measured observables,  are used  to calibrate the RMF model.

{\it Results and discussion:}
In the following, the results shown in Fig. \ref{fig_rho_T_sets_of_3} are discussed. 
  For all Xe+Sn systems, the extracted temperatures
distributions are relatively wide, however, the average values are very close to the simplified ideal gas assumption \cite{albergo} used in previous analyses \cite{Qin2011,indra,Pais2020prl}, see Fig. 6 Supplemental Material. Still, the temperature clearly increases as $v_{\text{surf}}$ increases, and a 
temperature evolution can be inferred from the data
like in the previous analyses.

  The same does not occur for the baryonic density: although a small density dependence can be identified with a slight decrease of the density when $v_{\text{surf}}$ increases, the results are also compatible with a single density, $\sim$0.015 fm$^{-3}$.
  The relatively large dispersion might signal that at each cooling step the different species decouple from slightly different points of the density profile, as expected in a dynamical process. However, within the explored density interval, the abundances vary only very slowly in the model, meaning that we can still consider  the  $v_{surf}$ sample as a statistical ensemble, and identify the average posterior density as
 the chemical freeze-out density 
  at the surface of the emitting source, 
  below which the particles do not feel anymore the nuclear interaction.

This result is at variance with previous analyses \cite{Qin2011,indra,Pais2020prl,Pais2020} where different prescriptions for the expected density were used, but the assumption of an effective density increasing with $v_{\text{surf}}$ 
 was always obtained in the adopted method of analysis. 
The direct extraction of the unknown effective density from the experimental data set via the Bayesian analysis is at the origin of the excellent reproduction of the experimental data in Fig. \ref{Mass_Fractions_vsurf_groups_of_three}. 
It might be interesting to observe that the freeze-out picture was already advanced to interpret INDRA vaporization data \cite{BORDERIE1996}, where multiplicities were very well reproduced with the explicit hypothesis that whatever the effective temperature, fragmentation data bear information on a single density, tentatively identified with the condition of chemical freeze-out.

 Taking these results into account, we will interpret the Bayesian results as follows: a) The selected INDRA data only give information about a single value of the baryonic density, which we will take to be $\approx$ 0.015~fm$^{-3}$, a density compatible with all systems. Note the correlation between $\rho$ and $T$, which introduces degeneracy in the determination of thermodynamic properties.  We expect that considering different systems will help to break this degeneracy. We will assume that the ratio $x_s$ does not vary with the baryonic density within the limited range explored by the data; b) The data test temperatures in the range $\sim 6$ MeV to $\sim 10$ MeV. There is a temperature dependence of the coupling $x_s$ corresponding to a weakening of the interaction with temperature, i.e. smaller values of $x_s$ correspond to the larger values of temperature.

\begin{figure}
	\centering\includegraphics[width=
 \linewidth,height=7cm]{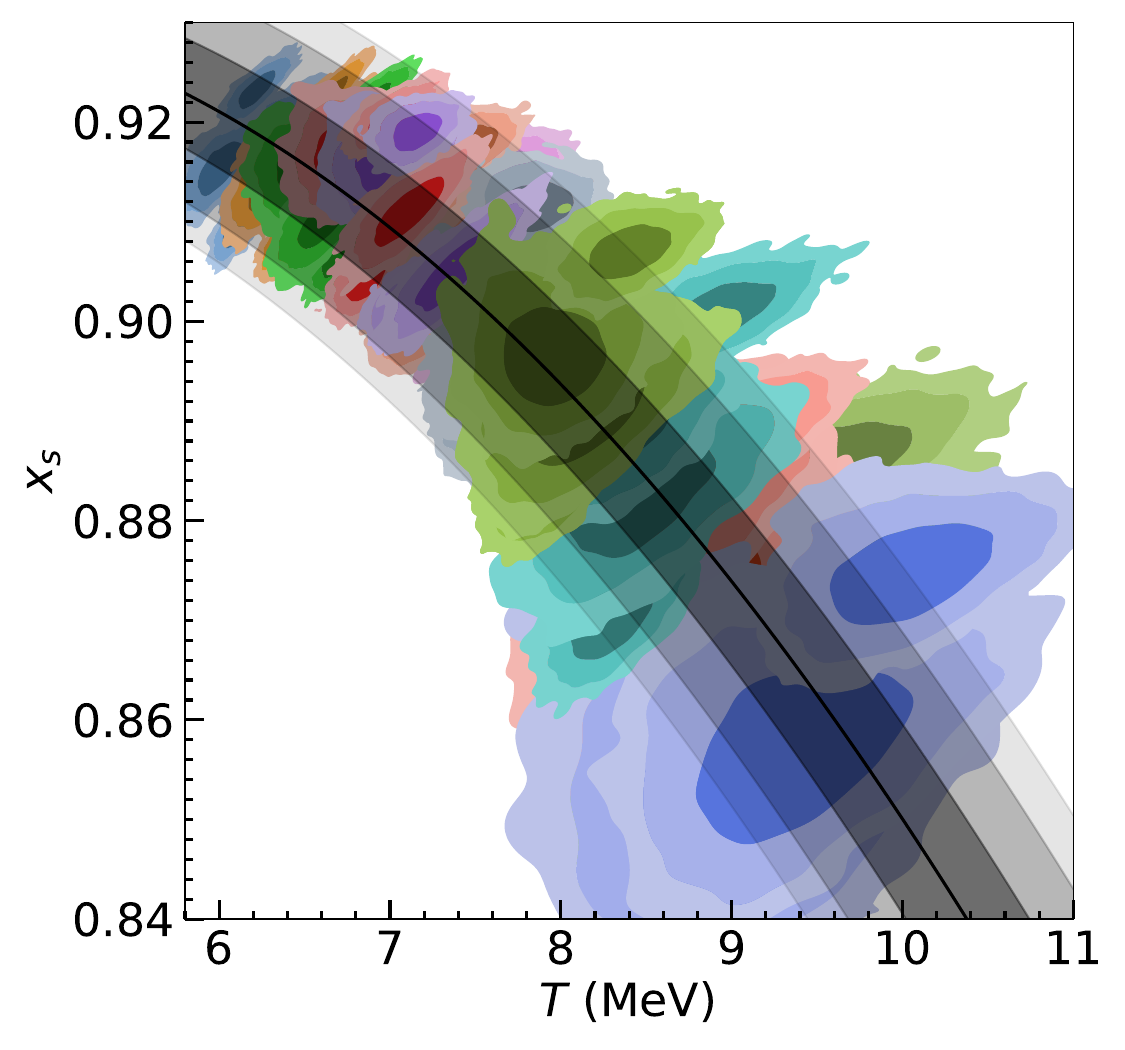} 
	\caption{\justifying  Posterior distributions of the effective cluster coupling parameter $x_s$ as a function of the temperature $T$ 
    for the FSU RMF model.
    Each colour corresponds to a specific $v_{\text{surf}}$ bin as in Fig. \ref{fig_rho_T_sets_of_3}) and the four distributions per colour correspond to the four independent inferences performed on the different colliding systems. A quadratic fit $x_s=aT^2+bT+c$ of the global data is represented in black, the median (black solid line) together with the 1,2,3-$\sigma$ uncertainty regions (the parameters of the fit are shown in Table \ref{table_parameters_xs_fit}) 
    for the FSU  and DD2 RMF models.
    }
 \label{fig_fix_dens_overlap_reg_Tvsxs_13}
\end{figure}

In Fig. \ref{fig_fix_dens_overlap_reg_Tvsxs_13} the calibrated values for the $x_s$ are plotted  as a function of the temperature, for all $v_{\text{surf}}$ values and colliding systems 
for the FSU RMF model (DD2 shows a similar behavior).
The decreasing trend of $x_s$ as a function of the temperature is clearly seen for all the four systems. 
As we can see from the figure, the deviation between 
the $x_s(\rho,T)$ obtained with the different entrance channels is smaller than the uncertainty of the parameter determination for each system. This 
  supports 
the statistical character of the samples.
Using the \texttt{lmfit} tools \cite{lmfit}, we performed a quadratic fitting of $x_s$ with respect to temperature $T$. The optimal quadratic model, $x_s = aT^2 + bT + c$, was chosen from several possibilities based on fit indices such as $\chi^2$, Akaike Information Criterion (AIC) and Bayesian Information Criterion (BIC) \cite{akaike1974new, schwarz1978estimating}. Notably, the BIC includes a penalty for models with more parameters. We consider the data of the four colliding systems together, see Table \ref{table_parameters_xs_fit} for the fit parameters along with 1$\sigma$ and 2$\sigma$ uncertainty.

The decreasing trend of $x_s$ as a function of $T$ is well described by this fit within 3-$\sigma$ confidence interval.

\begin{table}[h]
\centering
		\caption{\justifying Parameter estimates $a, b, c$ for FSU 
        and DD2
        with 1, 2$\sigma$ uncertainties for the model $x_s = aT^2 + bT + c$, as depicted in Figure \ref{fig_fix_dens_overlap_reg_Tvsxs_13}, where $T$ is measured in MeV.}
		\label{table_parameters_xs_fit}
  \setlength{\tabcolsep}{5.5pt}
      \renewcommand{\arraystretch}{1.5}
\begin{tabular}{ccccc}
\hline \hline
Parameter & Unit & Median & $1\sigma$ & $2\sigma$  \\ \hline
\hspace{-0.74cm}FSU $a$ & MeV$^{-2}$ & $-0.00203$ & $\pm{0.00003}$ & $\pm{0.00006}$ \\
$b$ & MeV$^{-1}$ & $0.01477$ & $\pm{0.00047}$ & $\pm{0.00093}$ \\
$c$ & & $0.90560$ & $\pm{0.0018}$ & $\pm{0.00355}$ \\
\\
\hspace{-0.74cm}DD2 $a$ & MeV$^{-2}$ & $-0.00171$ & $\pm{0.00003}$ & $\pm{0.00005}$ \\
$b$ & MeV$^{-1}$ & $0.01289$ & $\pm{0.00039}$ & $\pm{0.00077}$ \\
$c$ & & $0.91893$ & $\pm{0.00150}$ & $\pm{0.00296}$ \\
\hline

\end{tabular}
\end{table}

\begin{figure}
	\centering\includegraphics[width=
 \linewidth,height=7cm]{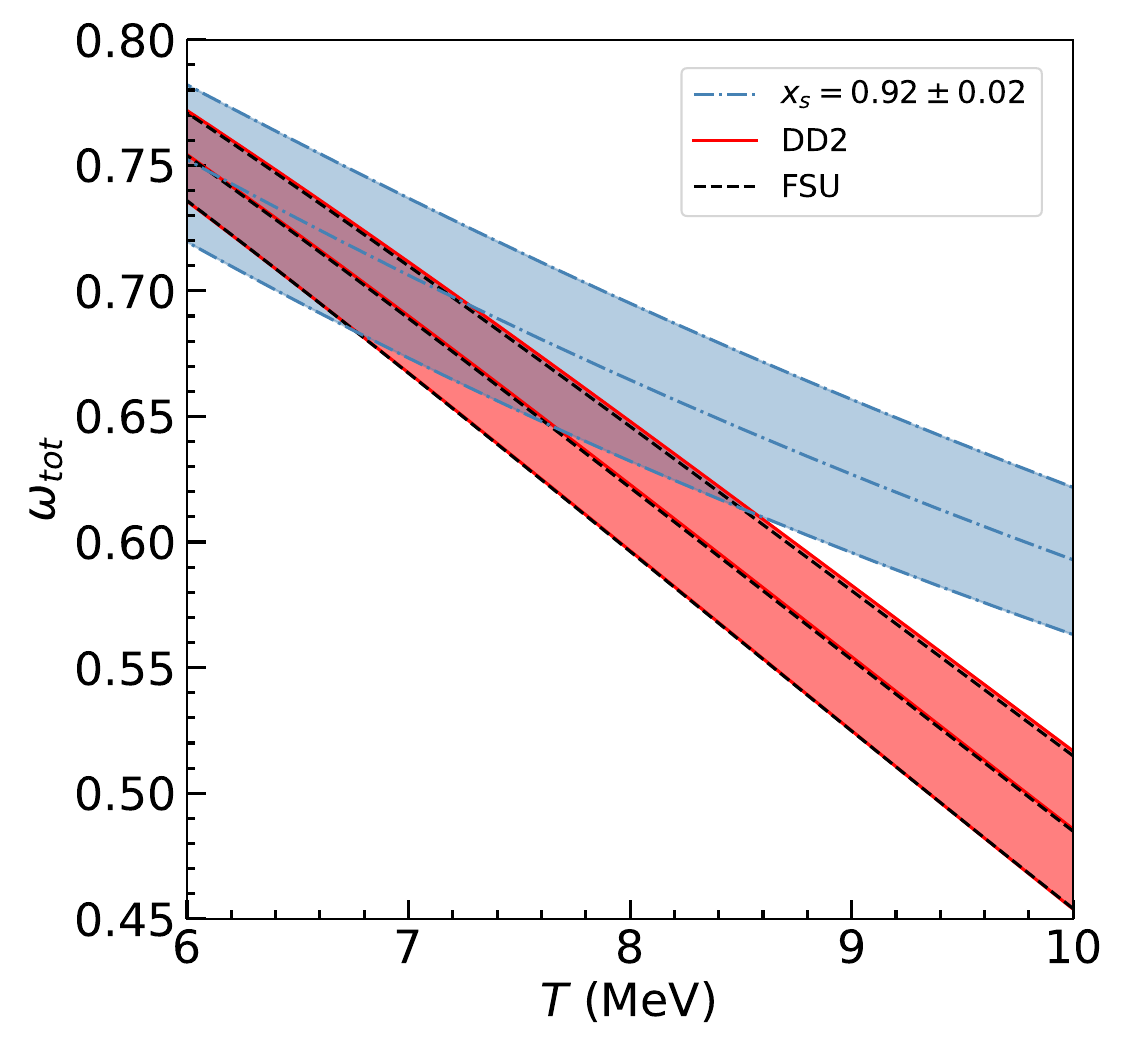} 
	\caption{\justifying  
    Total mass fraction predictions obtained
with the models  DD2 (filled red band with  red solid line contours) and FSU (unfilled band with
black dashed contours) as a function of  temperature $T$ with proton fraction
$y_q=0.45$ and baryonic density $\rho=0.015$ fm$^{-3}$  considering for
the $\sigma$-cluster coupling ratio $x_s$  the function $x_s(T)$
introduced in the present work. Both the  median and the  2-$\sigma$
uncertainties are plotted.  These results are compared to the mass
fractions determined with FSU and a temperature-independent value 
$x_s=0.92\pm0.02$ from Refs.\cite{Pais2020,Pais2020prl} (blue band
with dash-dotted delimiters).}
    
 \label{fig_Mass_Fractions_groups_of_3_vs_FSU_PRL_2020}
\end{figure}

In order to analyse how the temperature dependence of $x_s$ influences the light cluster abundances, in Fig. \ref{fig_Mass_Fractions_groups_of_3_vs_FSU_PRL_2020} we compare the total cluster abundances predicted by the RMF FSU 
and DD2 models
when considering this temperature dependence (red band), and when taking a constant value $x_s=0.92\pm0.02$ as in Refs.\cite{Pais2020,Pais2020prl} (blue band). A total proton fraction of 0.45 was considered as an example within the range of values explored in the four systems \cite{indra}.  Below $T\lesssim 8$ MeV the two bands overlap, but above this temperature the abundances predicted by the present study are systematically lower than the predictions of \cite{Pais2020,Pais2020prl}. This difference is larger than 20\%  for $T\sim10$ MeV. 
Since the $\sigma$-meson is responsible for the attractive strong force, a smaller $x_s$ corresponds to a weaker  cluster-$\sigma$ coupling, resulting in less bound clusters and, consequently, smaller abundances.  

{\it Conclusions:} 
 $^{136,124}$Xe$+^{124,112}$Sn central collisions at $32$MeV/nucleon  detected with the INDRA apparatus~\cite{Bougault_2018}, and sorted in bins  
 of the average  Coulomb corrected radial velocity $v_{\text{surf}}$,
 have been studied with an agnostic Bayesian analysis without a-priori hypotheses on the effective temperature and density explored by the different data samples.  
A Bayesian inference was performed within a RMF description of clusterized matter in order to determine the a-priori density and temperature dependent 
effective cluster couplings. 
We limit ourselves to  nuclear species for which the samples correspond to chemical equilibrium and finite size effects can be neglected.
The validity of the statistical treatment of the samples was a-posteriori 
  strengthened
by the fact that (i) excellent quality fits were obtained for all cluster species whose abundances verify chemical equilibrium \cite{Rebillard-Soulié_2024}, (ii) 
the
distributions 
obtained for the effective coupling ratio $x_s(\rho,T)$ using the four different entrance channels  
 do not show a discrepancy at the $3-\sigma$ level. 
The data showed a clear temperature dependence for $x_s$, but the density resolution was not sufficient to extract a possible density dependence (see also Fig 5 of Supplemental material). The temperature dependence was parametrized in terms of a quadratic function. The average density extracted from the data analysis, $\sim 0.015$~fm$^{-3}$ was interpreted as a freeze-out density.

 It is important to identify the two main differences considered in the present data analysis with respect to previous ones: (i) instead of calibrating the nuclear model using the equilibrium constants, which are ratios of the measured multiplicities that erase a large part of the entrance channel dependence and the associated information, 
in the present study the in-medium corrections are calibrated directly using the measured 
 particle abundances; this additionnally allows an a-posteriori verification of the equilibrium hypothesis by comparing the different entrance channels; (ii) in the previous analyses, the  
 effective temperature and baryon densities associated to the data samples  
 were estimated from the data within the hypothesis of a grand-canonical 
 ideal gas of classical clusters\cite{Qin2011,indra}. Possible effects of the in-medium corrections and quantum effects in the parameter estimation were already considered in \cite{Pais2020prl}, but only through a parametrized analytical correction to the ideal gas. In the present analysis, these hypotheses are relaxed and each estimation is performed within an independent Bayesian inference.  In this way, the only residual hypotheses concern the nuclear model in itself. 
 For this study, 
 employing the FSU and DD2 models, we have used an RMF quasi-particle approach, which has been shown in \cite{Pais2018} to be compatible with the quantum statistical description discussed in \cite{Ropke2015} and the generalized relativistic-density functional introduced in \cite{Typel2009}, 
 but it will be important to extend the Bayesian analysis also to the parameters of the nucleonic model in the future.

 Compared to previous results, the present study predicts a weaker attractive interaction at higher temperatures and, as a consequence, a faster dissolution of the clusters with temperature.  
 Our results have clear implications for environments such as a supernova core collapse or a binary neutron star merger, where the presence of clusters affects the transport properties.  In the future, it would be very interesting to be able to test different HI reaction mechanisms and entrance channels such as possibly exploring a wider range of temperatures and densities. Probably only a complete transport description of the HIC would lift the degeneracies found during the present study.

 \bigskip

 \bigskip
 \bigskip

 \bigskip
 \bigskip

 \bigskip
 \bigskip

 \bigskip
 \bigskip

 \bigskip
 \bigskip

 \bigskip
 \bigskip

 \bigskip
 \bigskip

 \bigskip

\section{Acknowledgements} 

This work was partially supported by  the IN2P3 Master Project NewMAC, the ANR project `Gravitational waves from hot neutron stars and properties of ultra-dense matter' (GW-HNS, ANR-22-CE31-0001-01), national funds from FCT (Fundação para a Ciência e a Tecnologia, I.P, Portugal) under projects 
UIDB/04564/2020 and UIDP/04564/2020, with DOI identifiers 10.54499/UIDB/04564/2020 and 10.54499/UIDP/04564/2020, respectively, and the project 2022.06460.PTDC with the associated DOI identifier 10.54499/2022.06460.PTDC. T.C. acknowledges the grant PRT/BD/154193/2022
(FCT, Portugal). H.P. acknowledges the grant 2022.03966.CEECIND (FCT, Portugal) with DOI identifier 10.54499/2022.03966.CEECIND/CP1714/CT0004. The authors acknowledge the Laboratory for Advanced Computing at the University of Coimbra for providing {HPC} resources that have contributed to the research results reported within this paper, URL: \hyperlink{https://www.uc.pt/lca}{https://www.uc.pt/lca}. We acknowledge support from Région Normandie under RIN/FIDNEOS and
support from Portugal/France PESSOA program (PHC 47833UB France,
2021.09262.CBM Portugal).

\newpage
\onecolumngrid
\appendix*

\bigskip
\begin{center}
   {\bf \large 
   Supplemental Material }
\end{center}

\twocolumngrid

\begin{figure}
		\includegraphics[width=0.9\linewidth,height=7cm]{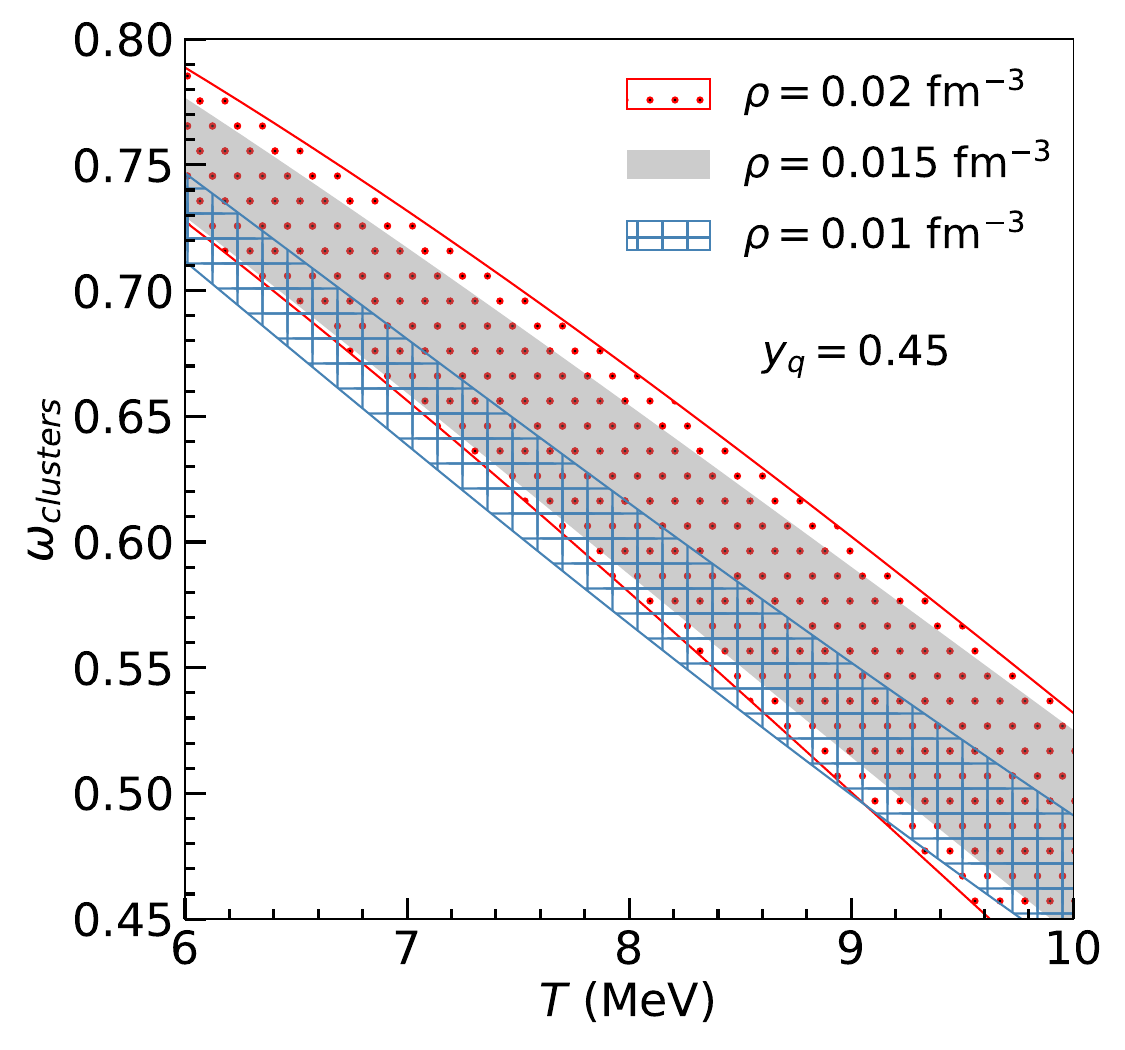}
	\caption{\justifying  Sum of the mass fractions of ${}^{2}\text{H}$,${}^{3}\text{H}$, ${}^{3}\text{He}$,${}^{4}\text{He}$ ($\omega_{clusters}$) for the FSU RMF model considering the quadratic dependence of $x_s$ with $T$, obtained for three different baryonic densities, 0.01, 0.015 and 0.02 fm$^{-3}$, and the proton fraction 0.45. The bands correspond to 3-$\sigma$ error bands of the quadratic fit. 
 }	
\label{fig_Mass_Fractions_Y_vs_T_vary_rho_fix_yq_groups_of_3_alternative}

\vspace{-0.23cm}

\end{figure}

\begin{figure}
	\centering\includegraphics[width=
 \linewidth]{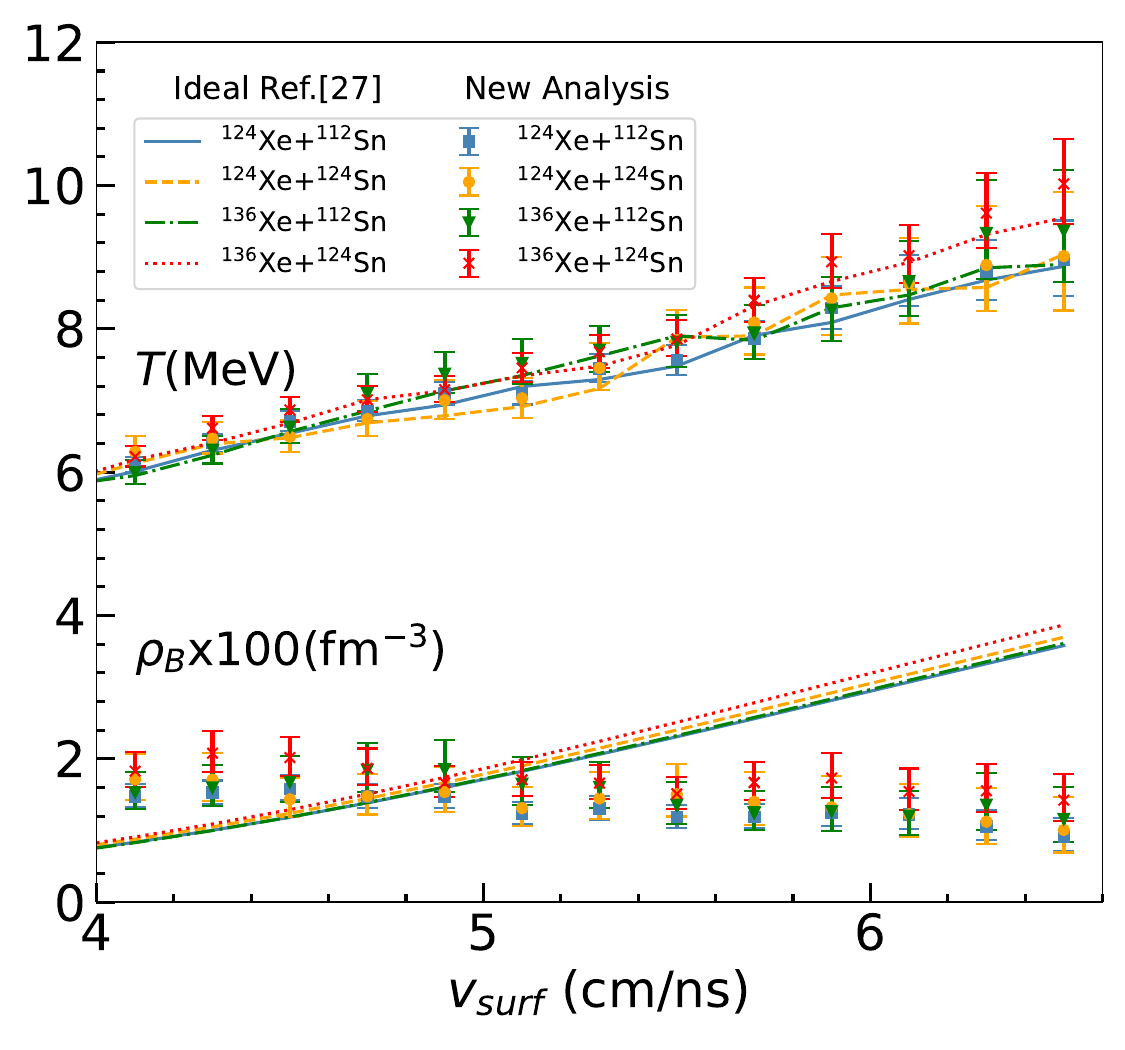} 
	\caption{\justifying Comparison between the calibrated values of the temperature and baryonic density (symbols with errorbars) performed within the bayesian framework used in the present work  for the FSU RMF model and the ones estimated considering the ideal gas assumption from Refs.\cite{indra} (lines).
} \label{T_rho_vsurf_ideal_gas_remi_comparison_sets_3}
\vspace{-0.23cm}
\end{figure}

To check the validity of the assumption that all $v_{\rm surf}$ measurements can be assigned to the same baryonic density, we plot in Fig. \ref{fig_Mass_Fractions_Y_vs_T_vary_rho_fix_yq_groups_of_3_alternative}, for the proton fraction 0.45, the abundance of light clusters for three different values of the baryonic density, 0.01, 0.015 and 0.02 fm$^{-3}$, which is the range of calibrated values shown in Figure 2 of the main letter.  The differences obtained are all within the error bands, supporting our assumption that there is no density dependence on $x_s$.

In Fig. \ref{T_rho_vsurf_ideal_gas_remi_comparison_sets_3}, we show the comparison between the calibrated values of the temperature and baryonic density  performed within the Bayesian framework used in the present work (symbols with errorbars)  for the FSU RMF model and the ones estimated considering the ideal gas assumption from Ref.\cite{indra} (lines). The temperatures in the ideal gas framework were obtained considering the statistical approach of Ref.\cite{albergo}. In both pictures, the results for the temperature are in agreement. The same does not occur for the baryonic density: in the ideal gas assumption, the density increases as a function of the $v_{\text{vsurf}}$; whereas the present results are compatible with a single density, $\sim$0.015 fm$^{-3}$ that we identify with the chemical freeze-out density, below which the particles do not feel anymore the nuclear interaction.  As observed in Figure 2 of the main Letter, the posterior distributions of $\rho$ and $T$ do not depend on the model and, therefore,  we obtain similar results for the DD2 RMF model.

\begin{figure*}
	\centering
		\includegraphics[width=0.8\linewidth]{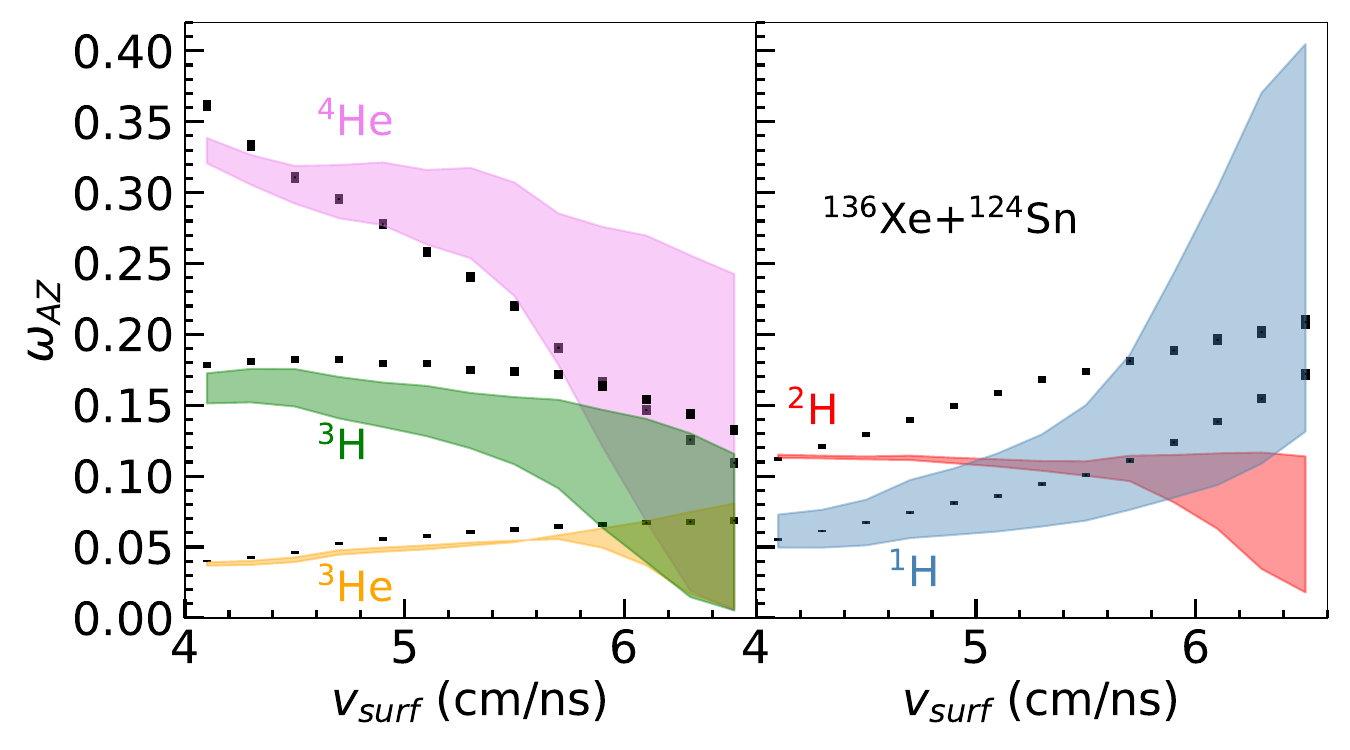}
	\caption{\justifying Comparison between experimentally measured mass fractions (black) \cite{indra} analysed under the assumptions of \cite{Pais2020prl}, and the mass fractions calculated with FSU RMF model using the temperatures, densities, proton fractions and $x_s$ obtained in \cite{Pais2020prl,Pais2020} (colour) for the system $^{136}$Xe+$^{124}$Sn. Note that $^6$He was included in the analyzis, and, therefore, the experimental data used differ from the one considered in the present study. 
 All quantities are represented with 1-$\sigma$ uncertainty. 
  }	
	\label{Mass_Fractions_vsurf_1system_Helena_data_136Xe_124Sn_PRL_2020_rho_T_then_obtained}
\end{figure*}

In Fig. \ref{Mass_Fractions_vsurf_1system_Helena_data_136Xe_124Sn_PRL_2020_rho_T_then_obtained}, we show the mass fractions experimentally measured in Refs.~\cite{indra,Rebillard-Soulié_2024} and analysed considering in-medium effects Refs.~\cite{Pais2020prl,Pais2020} as a function of the density, and we compare them with the theoretical ones calculated with the FSU model considering the extracted values of $\rho$, $T$ and $y_p$ from Refs.~\cite{Pais2020prl,Pais2020}, together with the $x_s$ coupling that was fitted to the chemical equilibrium constants found in that previous study. We notice that the deuteron, $^3$H and $^3$He abundances are not reproduced, and the experimental $^4$He abundances are outside the theoretical predictions for some temperature ranges, while the proton is reasonably well reproduced. 

\newpage

\bibliographystyle{apsrev4-2}
%

\end{document}